\documentstyle[aps,epsfig]{revtex}

\newcommand{\chapter}{\section}

\def\cm#1{}

\newcommand{\bfOmega}{{\Omega}}

\newcommand{\N}{N_c}
\newcommand{\dslash}{\partial\!\!\!/}

\newcommand{\gd}[1]{\gamma_{#1}}  
\newcommand{\ld}[1]{\lambda_{#1}}  
\newcommand{\p}{\partial}          
\newcommand{\f}[2]{\frac{#1}{#2}}
\newcommand{\tr}{{\rm tr}}
\newcommand{\be}{\begin{equation}}
\newcommand{\ee}{\end{equation}}
\newcommand{\beqn}{\begin{eqnarray}}
\newcommand{\eeqn}{\end{eqnarray}}
\newcommand{\Tr}{{\rm Tr}}

\newcommand{\s}{\sigma}

\newcommand{\Sig}{{\rho_0}}
\newcommand{\SigM}{{\Sig}}

\newcommand{\G}{\Gamma}

\newcommand{\calL}{\mbox{${\cal L}$}}
\newcommand{\calZ}{\mbox{${\cal Z}$}}
\newcommand{\calD}{\mbox{${\cal D}$}}
\newcommand{\psibar}{\bar{\psi}}
\newcommand{\etabar}{\bar{\eta}}

\newcommand{\betacrit}{\beta^{\rm cr}}

\renewcommand{\xi}{M}

\title{No Spontaneous Breakdown of Chiral Symmetry\\
 in Nambu-Jona-Lasinio Model}

\author{H. Kleinert \and B. Van den Bossche\thanks{Alexander von Humboldt Fellow,
on leave from absence of
 Physique Nucl\'eaire Th\'eorique, B5, Universit\'e de Li\`ege Sart-Tilman,
 4000 Li\`ege, Belgium}}

\address{Institut f\"ur Theoretische Physik, Arnimallee 14 D-14195 Berlin,
 Germany}

\begin{document}

\maketitle

\begin{abstract}
We argue that the spontaneous breakdown of  symmetry
in the chirally symmetric Nambu--Jona-Lasinio model which
was supposed to illustrate the origin of the low mass of pions in hadron
physics does
not occur due to strong fluctuations
in the $\s $ - $\pi$ field space. Although quarks acquire a constituent mass,
$ \s $  and $ \pi $ turn out to have equal
heavy masses of the order of the constituent quark
mass.
\end{abstract}

\chapter{Introduction}

The chirally symmetric Nambu--Jona-Lasinio model \cite{NJLM}
 was the first theoretical laboratory to illustrate
how light pions arise from a spontaneous breakdown of
chiral symmetry
in hadron physics. The
first realistic formulation
of the model which included flavored quarks,
possessed chiral symmetry $SU(3) \times SU(3)$,
and a spectrum of $\s, \pi, \rho, A_1$ mesons and their $SU(3)$
partners, was formulated and investigated in 1976 by one of the
 authors~\cite{hadroniz}, and has been the source of inspiration for many
 papers in nuclear physics in the past twenty years~\cite{vdb}.
 By eliminating the
Fermi fields in favor of a pair of collective
scalar and pseudoscalar  fields $ \s $ and $\pi$,
 as well as vector and axial vector mesons,
a Ginzburg Landau like collective field
action
was derived. This had been studied in detail earlier as an effective action
guaranteeing all
 low-energy properties
of hadronic strong interactions  which were known
from current algebra and partial conservation of
the axial current (PCAC).

In two important respects, however, the model was unsatisfactory.
First it was not renormalizable in four dimensions, but required a
momentum space cutoff  $ \Lambda $ to
 produce finite results. Moreover, to obtain
physical quantities of the correct size, the cutoff
had  to be rather small, below one~GeV, thus limiting the reliability of the
predictions to very low energies. Second, if the
fermions were identified with quarks, the model could not
account for their confinement.

The nonrenormalizability
was removed in~\cite{hadroniz} by
replacing the four-fermion interaction by the exchange
of a massive vector meson. The different attractive meson channels were
 obtained by a
Fierz transformation. The mass
of the vector meson took over the role of the cutoff.
The energy range of applicability was, however, not increased since the model
would still allow for free massive quarks.

The purpose of this note is to point out a much more severe
problem with the model which seems to invalidate most conclusions
derived from it in the literature: If chiral fluctuations are
taken into account in a certain nonperturbative approximation, the
spontaneous symmetry breakdown disappears, and the zero-mass pions
acquire the same mass as the $\s$-mesons, both of the same order
as the constituent quark mass. The nonperturbative nature of the
argument seems to be the reason why the phenomena has been
overlooked until now.

Since the Nambu--Jona-Lasinio model is incapable of accounting
for confinement,
it gave no reason
 for introducing colored
quarks. It is  curious to observe that the restoration of symmetry
by chiral fluctuations would offer such a reason, albeit with an
unphysical number of colors: the  physically desired spontaneous
symmetry breakdown conclusion can only be achieved by introducing
at least five identical replica of fermions. The existing three
colors are insufficient to save the purpose of the model.

 The non-perturbative arguments used
in this paper are analogous to those applied before
 in a discussion of the
Gross-Neveu model~\cite{GNM} in $2+\varepsilon$ dimensions
\cite{2phts}, where it was shown that this model has two phase
transitions, one where quarks become massive and another one where
chiral symmetry breaks spontaneously.
They have also been applied
to explain the experimental observation of
two transitions in
high-$T_c$ superconductors \cite{KBA}, and
to show that directional fluctuations
in Ginzburg-Landau theories
with spontaneously broken O($N$) symmetry
disorder the system {\em before\/}
size fluctuations of the order field become relevant
\cite{klcrit}.

\chapter{Nambu--Jona-Lasinio model}

Let us briefly recall the relevant features of the
Nambu--Jona-Lasinio model for our considerations.
The model contains $N_f$  quark fields $\psi(x)$, one for each
flavor. Each of them may appear with $N_c$ colors, such that the total
number of quarks is
$N=N_f\times N_c$. Since
the fluctuation phenomenon to be discussed will be caused by the almost
massless modes, we may restrict ourselves to the almost massless
up and down quarks. We will comment later on the effect of the
heavier quarks.

The Lagrangian of the model is given by \cite{vdb}
\be
\calL=\psibar\left( i\dslash-m_0 \right) \psi+\f{g_0}{2N_c}\left[
\left( \psibar\psi \right)^2+\left( \psibar\ld{a}i\gd{5}\psi
\right)^2 \right],
 \label{NJLModel} \ee
where an implicit summation over $a=1,2,3$ is assumed.
A small diagonal quark mass matrix $m_0$
  breaks slightly the
$SU(2) \times SU (2)$ part of
the chiral symmetry which lifts the mass of the pion
to a small nonzero value.
We have omitted the flavor symmetric vector gluon exchange used in
Ref. \cite{hadroniz} which would have given rise, after a Fierz
transformation, to additional vector and axial vector
interactions, which would not influence
the chiral fluctuations to be investigated here.
Thus we use the original nonrenormalizable
interaction corresponding to an infinite vector gluon mass.
The coupling constant in (\ref{NJLModel}) is defined with  the
number of colors
$N_c$
 in the denominator,
 to allow for
 a finite  $N_c\rightarrow \infty$ limit of the model at a fixed $g_0$.
The $2\times 2$-dimensional matrices $ \ld{a}/2,$ with
$a=1,\dots,3$ generate the fundamental representation of flavor
$SU(2)$, and are normalized by $\tr (\ld{a}\ld{b})=2\delta_{ab}$.

Via a Hubbard-Stratonovich transformation, the
Lagrangian~(\ref{NJLModel}) is converted into a theory of
collective scalar and pseudoscalar fields $\s$ and $\pi_a$.
Defining the propagator in the presence of the meson fields
\begin{equation}
G \equiv \f{i}{i\dslash-m_0-\s-i\gd{5}\ld{a}\pi_a},
\label{@GG}
\end{equation}
and adding external quark sources
 $ \eta ,\bar  \eta $,
 one can integrate out the quark fields
from the corresponding Lagrangian.
%
%
Summing over colors,
 the  generating functional of the Green functions
takes the well-known form
\be
\calZ=\int \calD\s\calD\pi\exp\left\{i\N\left[
 -i\Tr'\ln iG^{-1}-\f{1}{2g_0}\int d^Dx\,\left(
\s^2+\pi_a^2
\right)
+i\N^{-1}\int d^Dxd^Dy\,\etabar G\eta \right]\right\}.
\label{@effp}
\ee
The term inside the brackets
is the collective field action ${\cal A}[\s,\pi]$,  whereas the symbol
 $\Tr'$ includes both, the functional spacetime ``index"
 $x$, and the internal trace over  spin and flavor indices:
$\Tr'\equiv \int d^Dx\,\tr_{\gamma}\tr_f$.

By extremizing
${\cal A}[ \s ,\pi]$ at zero sources $ \eta ,\bar  \eta $,
we obtain
the field equation for the collective field $(\s , \pi_a)$:
\be
\tr_{\gamma}\tr_f \left[ G(x,x) {1 \choose i\ld{a}\gd{5}} \right]
= \f{1}{g_0}{\s(x) \choose \pi_a(x)}.
\label{@}
\ee
For constant fields, this equation becomes a {\em gap equation\/}.
Its solutions will be marked by a superscript ``s" for
``stationary phase approximation''.
 From now on, unless explicitly stated,
we shall consider the model with zero mass, $m_0 = 0$. The
stationary pseudoscalar solutions $\pi_a^{\rm s}$ can always be
chosen to be
 vanishing,
 while
the scalar solutions
can  be $\s^{\rm s}=0$, or
  $\s^{\rm s}\equiv  \Sig $.
In the first case, the ground state is chirally symmetric, in
the second the symmetry is spontaneously broken. This is the state
of physical interest whose stability will now be discussed.

\section{Effective potential and gap equation}

In the limit  $\N\rightarrow\infty$, the generating functional is
given {\em exactly\/} by the extremal field configurations, which
will be parameterized as $(  \sigma ^{\rm s}(x),\pi^{\rm
s}_a(x))=( \Sig(x),\mathbf{0})$. The system has an effective
action per quark
\be
\f{\G(\Sig,\Psi,\bar{\Psi})}{\N}=-i\Tr'\ln iG_{\Sig}^{-1}-\f{1}{2g_0}
\int d^Dx\Sig^2+\f{1}{\N}\int d^Dx\bar{\Psi}_aiG_{\Sig}^{-1}\Psi_a,
\label{EffActionperQuark}
\ee
where $\Psi=iG_{\Sig}\eta$ is the expectation value
$\langle\psi\rangle$ of the quark field, and $G_{\Sig}$
 its propagator
\be
G_{\Sig}
= \f{i}{i\dslash-\Sig}  .
\label{@propa}\ee
This shows that the solution of the gap equation with $ \Sig\neq
0$ describes  quarks with a nonzero mass $M=\rho_0$, which has
been generated by the spontaneous symmetry breakdown, and is
referred to as the {\em constituent quark mass\/}. In the present
approximation of zero bare mass $m_0$, the constituent quark mass
is about equal to 300 MeV for up and down quarks (see the
discussion in Refs.~\cite{hadroniz,QM}). In either case, the Green function
(\ref{@GG}) in the stationary field is diagonal in flavor space.

%
In the absence of external quark sources, the ground state
expectation value of a fermion field is always zero, and the
expectation value $\Sig(x)$ is constant, so
that~(\ref{EffActionperQuark}) reduces into

\be
\f{\G(\rho)}{\N}=-i\Tr'\ln iG_{\rho}^{-1}-\f{1}{2g_0}\int d^Dx\rho^2,
\label{EffectivePotential}
\ee
where we have allowed the fields $\sigma$ and $\pi_a$ to be
nonextremal, defining $\s^2+\pi_a^2\equiv \rho^2$, and reserving the
notation $\rho^2_0$ for the extremum. This is determined, after a
Wick rotation to euclidean momenta $p_E$ with $p_0=ip_{E,0},$
$id^Dp\rightarrow -d^Dp_E,$ $p^2\rightarrow -p_E^2$, by the gap
equation
%
%
%
\be
\f{1}{g_0}=2\times2^{D/2}\int \f{d^Dp_E}{(2\pi)^D}\f{1}{p_E^2+ \Sig^2}.
\label{GapEquation2}
\ee
We have divided the two sides of the gap equation by a common
factor $\Sig$, since we want to study the spontaneously broken
phase.

The gap equation must be regularized, which may be done in many
ways. Here, we shall use two methods: analytic continuation in the
dimension $D$, and a cutoff~$\Lambda$ in momentum space. The
former is mathematically more elegant and has the advantage of
relating the properties in four dimensions to those
in $2+ \varepsilon $.  It has, however, some unphysical properties
which require special attention, as we shall see.
  Such problems are absent
in a cutoff regularization scheme, which  exhibits
clearly
 the
 physical
 divergences
caused by the infinite number of degrees of freedom
of the field system.
Factorizing the integral in (\ref{GapEquation2})
into direction and size of the momentum $p_E$, we
%
 bring the gap equation to the form
\be
\f{1}{g_0}
=2\Sig^{D-2}\f{\G(1-D/2)}{(2\pi)^{D/2}}.
\label{GapEquation}
\ee
%
Denoting by
 ${\bfOmega}$ the $D$-dimensional volume $\int d^Dx$,
the volume density
$v(\rho)\equiv-\G(\rho)/{\bfOmega}$ of the effective action
(\ref{EffectivePotential}) is the {\em effective potential\/}  per
quark.
Performing the internal traces, and  subtracting a divergent
constant term associated with the chirally symmetric state with
$\Sig=0$, we obtain the  {\em condensation energy\/} in euclidean
space:
\be
\Delta v(\rho)=\f{\N}{2}\left[
\f{1}{g_0}\rho^2-\rho^{D}\f{4}{D}\f{1}{(2\pi)^{D/2}}\G(1-D/2)
\right].\label{EffectivePotential2}
\ee
In an even number of dimensions $D$, both the gap equation (\ref{GapEquation})
and the effective potential (\ref{EffectivePotential2}) are
divergent, due to a pole in the factor $\G(1-D/2)$. Introducing the
diverging parameter $b_{\epsilon}=2\G(1-D/2)/[D(2\pi)^{D/2}]$,
%
%
we can rewrite the gap equation and
effective potential in the more compact form
as
\beqn
 \f{1}{g_0}&=&D\Sig^{D-2}b_{\epsilon},
\label{bepsgap}
\\
 \Delta v(\rho)&=&\f{\N}{2}\left[
\f{1}{g_0}\rho^2-2\rho^{D}b_{\epsilon}
\right].
\label{bepspotential}
\eeqn
In the more physical regularization with
 a cutoff $\Lambda$ in momentum space,
these expressions  look more complicated:
\beqn
 \f{1}{g_0}&=&\f{2}{(2\pi)^2}
\left[
\Lambda^2-\Sig^2\ln\left(
1+\f{\Lambda^2}{\Sig^2}
\right)
\right],
\label{lambdagap}
\\
 \Delta v(\rho)&=&\f{\N}{2}\left\{
\f{1}{g_0}\rho^2-\f{2}{(2\pi)^2}\left[
\f{\rho^2\Lambda^2}{2}
+\f{\Lambda^4}{2}\ln\left(
1+\f{\rho^2}{\Lambda^2}
\right)-\f{\rho^4}{2}\ln\left(
1+\f{\Lambda^2}{\rho^2}
\right)
\right]
\right\}.
\label{lambdapotential}
\eeqn

The results
(\ref{bepsgap}) and~(\ref{bepspotential})
 of the analytic regularization scheme
 can be mapped
roughly into the cutoff results (\ref{lambdagap})
and~(\ref{lambdapotential}) if we recall the
special property of dimensional regularization that all
integrals over pure momentum powers
vanish identically: $\int d^D\!k\,(k)^{\alpha}=0$
(Veltman's rule). Thus, arbitrary
 pure powers of the cutoff $ \Lambda ^ {\alpha+D} $
have no counterpart in dimensional regularization.
Only
logarithmic
 divergences can be related
to diverging pole terms
 $1/\epsilon\rightarrow\infty$ for $ \epsilon \rightarrow 0$.
It is therefore inconsistent
 to  relate
  $ \epsilon $ to $ \Lambda $ by setting
 $\Gamma(\epsilon/2-1)\approx
 \Lambda^2/\rho_0^2$,
 as  proposed by Krewald and Nakayama \cite{krew}. Only the
logarithmic divergence in (\ref{lambdagap})
can be mapped to the small-$ \epsilon $
divergence in (\ref{bepsgap}), setting
  $\Gamma(\epsilon/2-1)\approx -\ln(1+\Lambda^2/\rho_0^2)$.
  With their inconsistent identification,
 Krewald and
  Nakayama matched $ \Lambda $ by an $\epsilon>2$
which lies
in the wrong region $D<2$, the physically
  relevant range being $D\in(2+\epsilon,4-\epsilon)$.
  Note that the matching of the logarithm
 at the level of the effective
  potential leads to the properly matched gap
  equation, thus having
circumvented
 the unphysical properties of
 the analytic regularization.
The free use of this scheme
in  renormalizable field theories
relies on the
fact
that all infinities
are eventually absorbed in unobservable
bare quantities, such that the artificial zeros of
the integrals over pure powers of momenta cannot produce problems.
In nonrenormalizable theories, on the other hand,
only a cutoff (or a
related Pauli-Villars regularization)
is   physical, and analytic regularization
  must be treated with caution.
 This
  is seen even more dramatically in integrals
  which do not have logarithmic infinities. For example the condensation
  energy~(\ref{EffectivePotential2}) in $D=3$ dimensions
would be a finite negative number in analytic regularization,
 while being
   a linearly divergent positive function of the cutoff.

\section{Chiral fluctuations}

 Since the physical number of quarks
$\N$ is finite, the fields perform fluctuations of magnitude
 $1/\sqrt{\N}$
 around their extremal value.
As long as $\N$ can be considered as a large number,
the deviation from the extremal field
configuration
 %
$(  \s',\pi'_a)\equiv
(  \s- \Sig,\pi_a)$
%
are small, and the action can be expanded in powers
of $(  \s',\pi'_a)$.
The quadratic terms in this expansion
define the propagators of the collective
fields $( \s',\pi'_a)$.
The higher expansion terms of
the trace of the logarithm in (\ref{@effp})
define the interactions.
With this decomposition, the inverse of
the quark propagator~(\ref{@GG})
can be decomposed into  a constant and a fluctuating part, setting
%
$iG^{-1}=iG_{ \Sig}^{-1}-(\s'+i\gd{5}\ld{a}\pi'_a)$,
%
with $G_ {\Sig} $ of Eq.~(\ref{@propa}).
Then we have
\be
\Tr\ln iG^{-1}=\Tr\ln iG_\SigM ^{-1}+\Tr\ln\left[
1+iG_\SigM \left(
\s'+i\gd{5}\ld{a}\pi'_a
\right)
\right].
\ee
An expansion of the last term up to the second order in the fields
gives an approximate partition function (with
$\calZ_0\equiv\exp\left[-\bfOmega_E \N v(\SigM)\right]$ and $\bfOmega_E$ is
 the euclidean volume)

\be
\calZ=\calZ_0\int\calD\s\calD\pi\exp\bigglb(
i\N\left\{
\f{i}{2}\Tr'\left[
iG_\SigM \left(
\s'+i\gd{5}\ld{a}\pi'_a
\right)
\right]^2-\f{1}{2g_0}\int d^Dx({\s'}^2+{\pi'_a}^2)
\right\}
\biggrb).
\label{SecondOrderAction}
\ee
%
%
%
The functional matrix between the fields
in the exponent gives us
directly the
 inverse of the  desired collective free field  propagators $G_{\s'},
G_{\pi'}$. In momentum space, we identify
\be
{\cal A}_0[\s',\pi'] = \f{1}{2}\int d^Dq\left[
\pi'_a(q)G_{\pi}^{-1}\pi'_a(-q)
+\s'(q)G_{\s}^{-1}\s'(-q)
\right],
\label{@effa}\ee
where
%
%
%
%
%
%
%
%
%
\be
G_{\s,\pi}^{-1}=2\times2^{D/2}\N\int_0^1 dy\int
\f{d^Dp_E}{(2\pi)^D} \f{ q_E^2+p_Eq_E+(2\SigM^2,0)}{\left[
(q_E^2+2p_Eq_E)y+p_E^2+\SigM^2 \right]^2}.
\label{SigPropStiffLambdax} \ee
%
In this expression, the gap equation
(\ref{GapEquation2}) has been
 used to eliminate the term $1/g_0$.
The notation $(2\SigM^2,0)$ indicates
that only the equation for $\s$ contains an extra term $2\SigM^2$.

In four spacetime dimensions,
the integral evaluated in dimensional regularization reduces
 to $q_E^2/2$ for the pseudoscalars,
and to $(q_E^2+4\SigM^2)/2$ for the scalars,
both
 with
a diverging coefficient. The first leads to a
zero mass for pions as a manifestation of Goldstone's theorem, the
second to a mass equal to twice the constituent quark mass for the
$\sigma $-mesons. For a finite result, the integrals must
be regularized.
In $D=4- \epsilon$ dimensions,
the inverse
euclidean propagator is seen to start out for small $q^2_E$  like
%
%
%
\be
G_{\pi}^{-1}\approx\N\left(1-\f{D}{2}
\right)
Db_{\epsilon}\f{1}{\SigM^{4-D}}\f{q_E^2}{2}\equiv Z_{\pi}^{(\epsilon)}( \rho _0)q_E^2
+{\cal O}(q^4_E).
\label{PiPropStiffEps}
\ee
with the same
$b_{\epsilon}$ as defined above Eq.~(\ref{bepsgap}).
If the theory is regularized with a cutoff $\Lambda$
in momentum space,
this becomes
\be
G_{\pi}^{-1}=\f{\N}{(2\pi)^2}
\left[
\ln\left(
1+\f{\Lambda^2}{\SigM^2}
\right)
-\f{\Lambda^2}{\Lambda^2+\SigM^2}\right]
q_E^2\equiv Z_{\pi}^{(\Lambda)}( \rho _0)q_E^2.
\label{PiPropStiffLambda}
\ee
In the right-hand part of the two equations, the factors
 in front of $q^2_E$ have been identified as
the wave function
renormalization constants $Z_\pi( \rho _0)$ of the pion field
in the two regularization schemes.

As a consequence of the spontaneous symmetry breakdown, the
fluctuations of the pseudoscalar fields are massless. These fields
appear in the $x$-space version of the  action (\ref{@effa}) in  a
pure gradient form
\be
{\cal A}_0[\pi']= \f{\beta}{2}\int d^Dx\left\{ [\partial
\pi'_a(x)]^2 \right\}, \label{@prop}
\ee
with $\beta=Z_{\pi}$.
Due to chiral symmetry, this gradient action
can be
extended
to the
gradient action of
an arbitrary field $( \sigma ,\pi_a)$. Introducing
the directional unit vector fields
$n_i=(\hat{\s}',\hat{\pi}_a')\equiv(\s',\pi_a')/\rho$, we find:
\be
{\cal A}_0[n_i]=\f{\beta(\rho^2)}{2}\rho^2\int d^Dx\,(\p n_i)^2 \mbox{,
~~~~~$i=1,\dots,N_n$,} \label{NLSig}
\label{@22}\ee
with $N_n=4$ and
\begin{equation}
 \beta ( \rho )=Z_{\pi}( \rho ).
\label{@}\end{equation}
This chirally invariant
action describes the massless pions with {\em all}
multipion interactions.\footnote{Only two approximations are
involved: the first one consists in freezing the size $\rho$ of
the fluctuations. The second one neglects corrections due to the
finiteness of the sigma mass. The latter corrections are expected
to be of the order $f_{\pi}^2/(4M^2)\approx3\%$.} The prefactor
$\beta$ is called the {\em stiffness of the directional fluctuations\/}
\cite{2phts,klcrit,bkt,cmab,GFCM}. In analytic regularization,
the result~(\ref{PiPropStiffEps})
shows that the stiffness of pion fluctuations
in $D$=2 dimensions becomes
\be
\beta=\f{\N}{2\pi\rho^2_0},
\label{StiffGN}
\ee
thus coinciding with the stiffness calculated  in Ref.~\cite{2phts}
in the Gross-Neveu  model (which contained a factor $N$
to be identified with the present $N_f\times N_c=2N_c$).

With the more physical cutoff regularization in $D=4$ dimensions, the
stiffness of
 directional fluctuations is
\be
\beta=Z_{\pi}^{(\Lambda)}( \rho _0)=\f{\N}{(2\pi)^2}
\left\{
\ln\left[1+
\left(\f{\Lambda}{\SigM}\right)^2
\right]-\f{\Lambda^2}{\SigM^2+\Lambda^2}
\right\}.
\label{StiffFullCutOff}
\ee
%
%
%

This is the crucial quantity leading to our fatal conclusions for
the restoration of
chiral symmetry.
 The stiffness
(\ref{StiffFullCutOff}) is far too small to let the directional
field settle in a certain direction, required for spontaneous
symmetry breakdown. The disordering effect of phase fluctuations
is well-known from many model studies of the $0(4)$-symmetric
Heisenberg model on a lattice. High-temperature expansions and
Monte Carlo simulations have shown that there exists a critical
stiffness below which the system goes over into a disordered
state.

For an analytic estimate of the critical stiffness,
 we relax the unit vector constraint for the vectors $n_i$ in~(\ref{NLSig})
 by introducing an additional
field
$\lambda(x)$ playing the role of a Lagrange multiplier. The
$n_i$-fields can then be integrated out in the partition function,
 leading
to an action
\be
S=\f{N_n}{2}\Tr\ln\left[ -\p^2+\lambda(x)
\right]-\beta(\rho^2)\rho^2\int d^Dx \f{\lambda(x)}{2},
\label{Newaction} \ee
where $\Tr$ denotes the functional trace (the summation over the fields
component has already been performed). For a large number $N_n$ of components,
the fluctuations are suppressed, and the
field $\lambda(x)$ becomes a constant satisfying a second gap equation
\be
\beta=\f{N_n}{\rho^2}\int \f{d^Dk}{(2\pi)^D}\f{1}{k^2+\lambda}.
\label{@secge}
\ee
If there is a nonzero solution $ \lambda \neq 0$, this will play
the role of a square mass of the $n_i$-fluctuations, and represents an
order parameter in the directional phase transition. The model has
a phase transition at a  critical stiffness
\be
\beta_c=\f{N_n}{\rho^2}\int \f{d^Dk}{(2\pi)^D}\f{1}{k^2}.
\label{CriticalStiff}
\ee
For a smaller stiffness, the phase fluctuations are so violent that the system
goes into a disordered phase with $ \lambda \neq 0$
giving all fields $n_i$ a nonzero square mass
$\lambda$. Since the fields $n_i$ are the normalized $ \sigma $ and $ \pi_a $
fields of the model, this determines an
equal nonzero square mass of
$ \sigma $ and $ \pi_a $ mesons, and thus a restoration
of chiral symmetry.

Note that the quarks are still massive: their constituent mass is a
consequence of the {\em formation} of the pairs, which are
strongly bound for small $\N$. The phase transition taking place
at the critical value of the stiffness, on the other hand, is
related to the Bose-Einstein condensation of the pairs. At small
$\N$, the two processes are widely separated. This separation of
the two transitions (pair formation and pair condensation) can be judged by
 the simple fluctuation criterion in Ref.~\cite{klcrit}.

 In our model, the number $N_n$ is equal to four, which is not very
large. Fortunately, Monte Carlo
 studies of the model \cite{Lee,neuhaus,horgan} have shown that $N_n=4$
is large enough to ensure the existence of the transition and the
quantitative reliability of the theoretical estimate of the
critical stiffness~(\ref{CriticalStiff}). From an evaluation
of~(\ref{CriticalStiff}) {\em on a lattice\/},
 and a comparison with Monte
Carlo studies, we estimate that the critical stiffness obtained
from~(\ref{CriticalStiff}) is correct to within less than 2\%
\cite{montecarl,neuhaus} or 6\% \cite{montecarl,Lee,horgan}. The same maximal error
is expected
if we
work in the continuum
using a momentum cutoff scheme.

For $N_n=4$ and a  cutoff
$\Lambda_{\pi}$ in the integral
(\ref{CriticalStiff}) over pion momenta,
 the critical stiffness is given by
\be
\beta_c=\f{4}{16\pi^2}\f{\Lambda^2_{\pi}}{\rho^2}.
\label{StiffNLSigma}
\ee
By comparing this
with the stiffness of the model in
(\ref{StiffFullCutOff}), we find
\be
N_c=\left(\f{\Lambda_{\pi}}{\Lambda}\right)^2\left(\f{\Lambda}{\SigM}\right)^2
\left\{
\ln\left[
1+\left(
\f{\Lambda}{\SigM}
\right)^2
\right]-\f{\left(
\Lambda/\SigM
\right)^2}{1+\left(
\Lambda/\SigM
\right)^2}
\right\}^{-1}.
\label{StiffCond}
\ee
This equation determines the number $N_c$ of identical quarks
which is necessary to produce a large enough
stiffness $\beta$ to prevent
the restoration of chiral symmetry. Only if the number of colors
exceeds this critical value, will the model possess a
phase in which the pion is a massless Goldstone boson,
 and $\sigma $  a
meson with a mass twice as large as that of the
constituent quarks. The critical number (\ref{StiffCond}) is
plotted as the solid curve in
 Fig.~\ref{fig1} for
$\Lambda_{\pi}=\Lambda$. We see that
 $N_c=5$ would be the smallest allowed value.
This number, however, is incompatible with color SU(3). This
suggests that the Nambu--Jona-Lasinio
model
always remains in the
symmetric phase, due to chiral fluctuations. It can therefore
not be used to
describe the chiral symmetry breakdown of quark physics, as has
been claimed by many publications, which have appeared in particular in
nuclear physics~\cite{vdb}.

 Can this conclusion be avoided by a different choice of
 parameters?
 To obtain a critical value smaller
than $N_c=3$ would require a pionic cutoff
$\Lambda_{\pi}\lesssim0.8\Lambda$. However, the cutoff cannot be chosen at will.
Let us study the cutoff dependence more precisely.
For this, we
refine the previous crude estimate
(\ref{CriticalStiff}), (\ref{StiffNLSigma}) of the critical
stiffness, which will henceforth be called Approximation 1,
by taking better account of the shorter wavelength fluctuations,
replacing the action
(\ref{@22})  by
\begin{equation}
{\cal A}_1[n_i]=\frac{\rho ^2}{2}\int d^Dx
\,n_i(x)G^{-1}_\pi(-\partial^2 )n_i(x),
\label{@propp}
\end{equation}
with $G^{-1}_\pi(-\partial^2 )$ from
Eq.~(\ref{SigPropStiffLambdax}). This exchanges $1/k^2$ in
Eq.~(\ref{CriticalStiff}) by the full pion propagator $G_\pi(k^2)/
Z_{\pi}^{(\Lambda)}$ associated with the action (\ref{@propp}).
The cutoff $ \Lambda_\pi$ makes the integral over pion momenta
finite. Its size is fixed by physical considerations.
The pion fields in the symmetry-broken phase are
composite, and will certainly not be defined over length scales
much shorter than the inverse binding energy of the pair wave
function, which is equal to $2M=2\rho_0$.
Thus we perform
the integral in the modified equation (\ref{CriticalStiff}) up to the
cutoff $4M^2$. This is Approximation 2, yielding the solid curve in Fig.~\ref{fig3}.

The phase with broken symmetry for three colors would be reached only if
 the quark loop integration is cut off at $ \Lambda ^2\gtrsim11M^2$.
Such a large value, however, is incompatible with the experimental
value
 of the pion decay constant
$f_\pi\approx 0.093$ which is given, in the large-$N_c$ limit of the model,
by
$f_\pi/M=Z^{1/2}(M)$. For typical estimates
of constituent quark masses $M\in(300,400)$ MeV \cite{hadroniz},
we find that $ \Lambda ^2/M^2$ should lie
in the range $(3.3,7.3)$, the highest value corresponding to the lowest
possible mass 300 MeV.

The above study has given us only the critical point, where the
pion mass goes to zero. We can do more and determine the common
nonzero square masses $ m_ \sigma ^2=m_\pi^2=\lambda $ of $\sigma$ and $\pi_a$-fields in
the phase of restored chiral symmetry. This is the subject of
 the next section.

\section{Meson masses}
The chiral fluctuations give rise to a change
 of the effective potential.
 They add to $\Delta v(\rho)$ in  Eq.~(\ref{lambdapotential})
 an additional energy coming from the stationary point of the
 action~(\ref{Newaction})
  at a constant $\lambda(x)= \lambda $:
\beqn
\Delta'_1 v(\rho,\lambda)&=&-\f{1}{2}
\lambda Z_0\rho^2+\f{N_n}{2}
\int_0^{\Lambda_{\pi}^2}\frac{dq_E^2\,q_E^2}{16\pi^2}
\ln[q_E^2 + \lambda ],
\label{lambdapotential2approx1}\\
\Delta'_2 v(\rho,\lambda)&=&-\f{1}{2}
\lambda Z(\rho)\rho^2+\f{N_n}{2}
\int_0^{\Lambda_{\pi}^2}\frac{dq_E^2\,q_E^2}{16\pi^2}
\ln[G^{-1}(q_E^2)/Z(\rho) + \lambda ],
\label{lambdapotential2approx2}
\eeqn
for Apprs.~1 and~2, respectively,
where the latter has
 $- \partial ^2$ replaced by $ G^{-1}_\pi(-\partial^2 )/Z(
\rho )$.
Extremizing $\Delta v(\rho)+\Delta'_{(1,2)}
v(\rho,\lambda)$ yields two coupled gap equations replacing
the independent gap equations (\ref{lambdagap})
and (\ref{@secge}). Introducing the reduced quantities
 $\bar{Z}(x)=\ln\left(1+x^{-1}\right)-(1+x)^{-1}$, and
 $x\equiv\rho^2/\Lambda^2,
y\equiv\lambda/\Lambda^2$,  we have for
Appr.~1:
\beqn
x_0\ln\left(1+x_0^{-1}\right)+\f{y}{2}\f{d}{dx}\left[
x\bar{Z}(x)\right]&=&x\ln\left(1+x^{-1}\right),\label{xeqn}\\
N_cx\bar{Z}(x)&=&\left(\f{N_n}{4}\right)\left\{
\left(\f{\Lambda_{\pi}}{\Lambda}\right)^2
-y
\ln\left[1+\left(\f{\Lambda_{\pi}}{\Lambda}\right)^2
y^{-1}\right]\right\}
\label{yeqn}.
\eeqn
For Appr.~2,
 the coupled gap equations are more complicated since the full
$q^2$-dependence of $Z_{\pi}$ has to be taken into account. They
read
\beqn
&&\hspace{-2.5cm}x_0\ln\left(1+x_0^{-1}\right)+
\f{N_n}{8N_c}\left\{
\int_0^{(\Lambda_{\pi}/\Lambda)^2}
\!\!\!\!\!\!k^2dk^2 \left[\f{\bar{Z}(x_0)}{\bar{Z}(k^2,x_0)}\right]
\f{d}{d x_0}\left[\f{\bar{Z}(k^2,x_0)}{\bar{Z}(x_0)}
\right]
\right\}
+\f{y}{2}\f{d}{dx}\left[
x\bar{Z}(x)\right]\nonumber\\
&=&x\ln\left(1+x^{-1}\right)
-\f{N_n}{8N_c}\left\{
\int_0^{(\Lambda_{\pi}/\Lambda)^2}
\!\!\!\!\!\!
\f{k^4dk^2}{k^2\left[\bar{Z}(k^2,x)/\bar{Z}(x)\right]+y}
\f{d}{d x}\left[\f{\bar{Z}(k^2,x)}{\bar{Z}(x)}
\right]
\right\}
,\label{xeqnapprox2}\\
N_cx\bar{Z}(x)&=&\left(\f{N_n}{4}\right)
\int_0^{(\Lambda_{\pi}/\Lambda)^2}
\!\!\!\!\!\!\f{k^2dk^2}{k^2\left[\bar{Z}(k^2,x)/\bar{Z}(x)\right]+y}
\label{yeqnapprox2},
\eeqn
where $\bar{Z}(k^2,x)$ is taken from the pion
 propagator~(\ref{SigPropStiffLambdax}), and is given by
\be
\bar{Z}(k^2,x)=\left(\f{N_c}{2\pi^2}\right)
\int_0^1p^2dp^2\int_0^1\f{(1-z)dz}{\left[p^2+k^2z(1-z)+x\right]^2}.
\ee
 There is no need to  write down the lengthy  analytic
solution to this double integral.
 The coupling constant $g_0$ has been eliminated in favor of
 the reduced mass $x_0$  which characterizes
the model uniquely above $N_c^{\rm cr}$, where $y=0$ and thus
 $\lambda=0$.
In that case, Eq.~(\ref{xeqn}) of Appr.~1 reduces
to~(\ref{lambdagap}).
 Equations~(\ref{yeqn}) and~(\ref{yeqnapprox2}), on the other hand,  determine
 the common square mass $ \lambda $ of $\sigma$ and $\pi_a$ as a
function of $N_c$, which  begins developing for $N_c < N_c^{\rm
cr}$. Note that going from Appr.~2 to 1
 corresponds
to using a momentum-independent pion normalization
$\bar{Z}(k^2,x)=\bar{Z}(x)$. This makes
 (\ref{xeqnapprox2}) and~(\ref{yeqnapprox2}) coincide with (\ref{xeqn}) and~(\ref{yeqn}).

\begin{figure}[h]
\unitlength 1cm
\begin{center}
\begin{picture}(10,8)
\put(0,0){\vbox{\begin{center}
\psfig{file=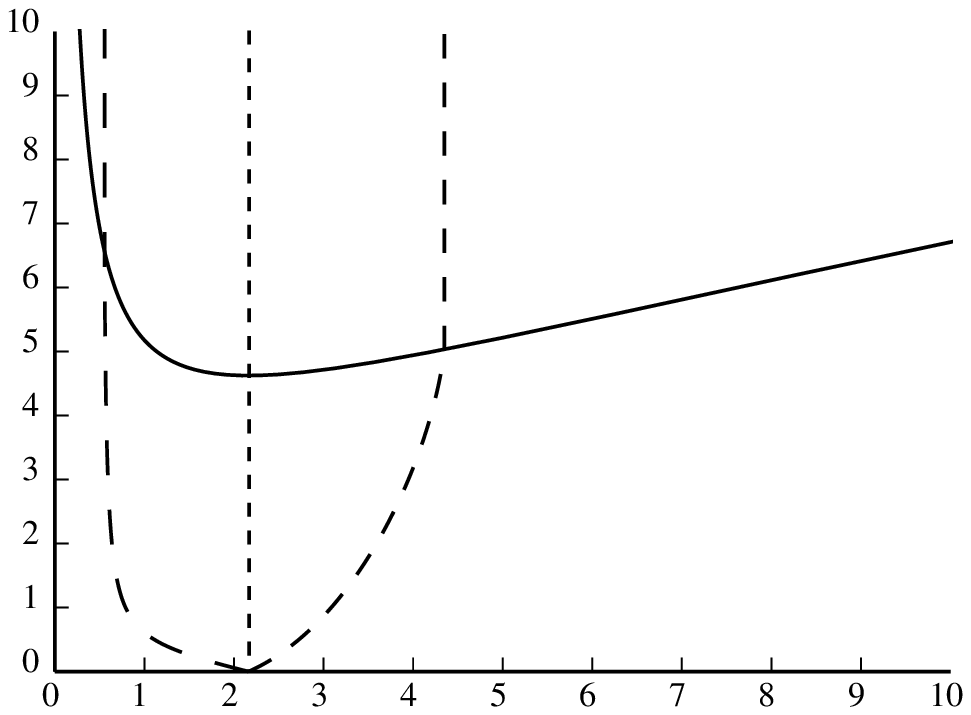,width=9cm}\end{center}}}
\put(0.5,3.2){\footnotesize $N_c$}
\put(4.5,0.15){\footnotesize $(\Lambda/\rho)^2$}
\put(3.05,5.5){\footnotesize$m_{\pi}=0$}
\put(3.05,5){\footnotesize$m_{\sigma}=2\rho_0$}
\put(5.2,2.75){\footnotesize$m_{\pi}^2=m_{\sigma}^2=\lambda\ne0$}
\put(6.5,4.1){\footnotesize $N_c^{\rm cr}$}
\end{picture}
\end{center}
\caption[]{Solid curve shows
{\em Approximation~1}  to critical number of colors $N_c^{\rm cr}$
as a function of the extremal value of
  $\rho=\sqrt{\s^2+\pi_a^2}$,
 above which chiral symmetry is restored.
 The dashed curves indicate the solutions
to the two gap equations
(\ref{xeqn}) and (\ref{yeqn}) for
three different values of the constituent quark mass
 $\rho_0=M$ in the symmetry-broken phase above $N_c^{\rm cr}$.
 The three quark masses
lie
above ($\SigM>\rho^*$), below ($\SigM<\rho^*$),
and at
 $\rho^*=M^*\approx \sqrt{0.46} \Lambda$
where $N_c^{\rm
cr}$ takes the minimal value 4.62,
with a constituent quark mass above $N_c^{\rm cr}$
of $ 0.678\Lambda$ (short-dashed curve). The medium-dashed  curve
corresponds to a constituent quark mass $0.479\Lambda$, and the
long-dashed to $ 1.342\Lambda$.} \label{fig1}
\end{figure}

\begin{figure}[h]
\unitlength 1cm
\begin{center}
\begin{picture}(10,8)
\put(0,0){\vbox{\begin{center}
\psfig{file=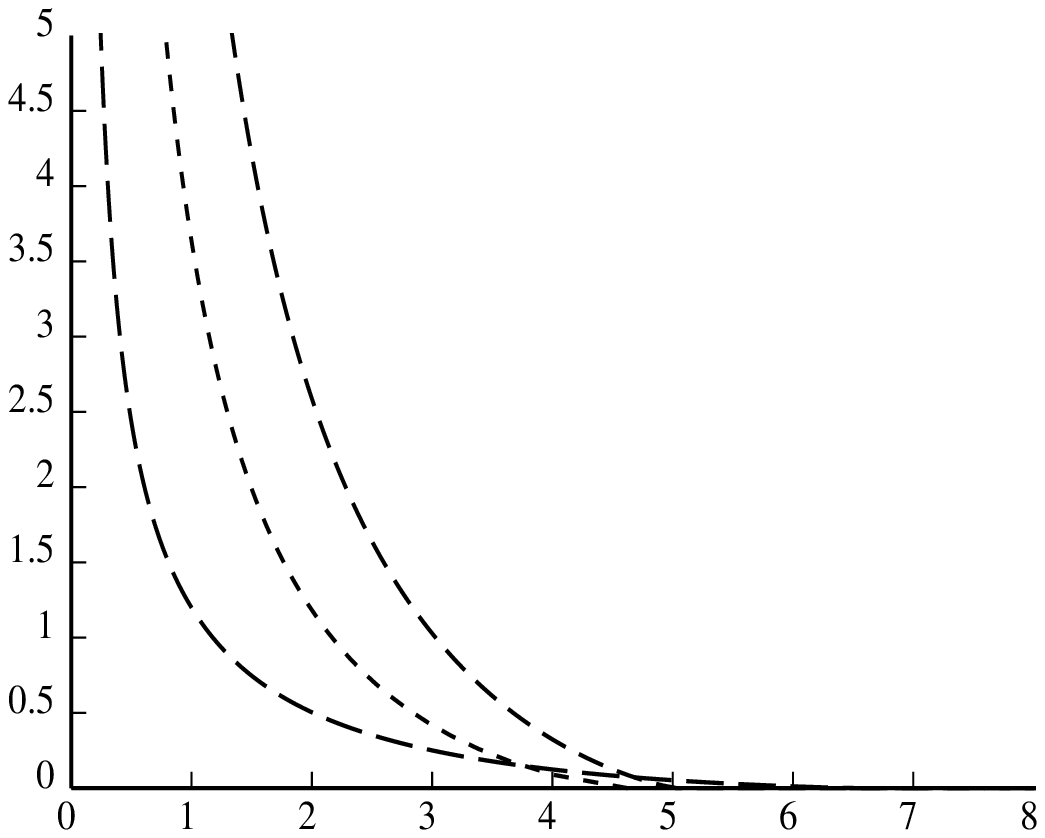,width=9cm}\end{center}}}
\put(0.5,3.5){\footnotesize $\frac{\lambda}{\rho_0^2}$}
\put(4.5,0.15){\footnotesize $N_c$}
\end{picture}
\end{center}
\caption[]{Common square masses
 $m_{\s}^2=m_{\pi}^2=\lambda$ as a function of $N_c$
in {\em Approximation~1}.
The three curves start at different critical values $N_c^{\rm cr}$
which can be read off Fig.~\ref{fig1}.}
\label{fig2}
\end{figure}

\begin{figure}[h]
\unitlength 1cm
\begin{center}
\begin{picture}(10,8)
\put(0,0){\vbox{\begin{center}
\psfig{file=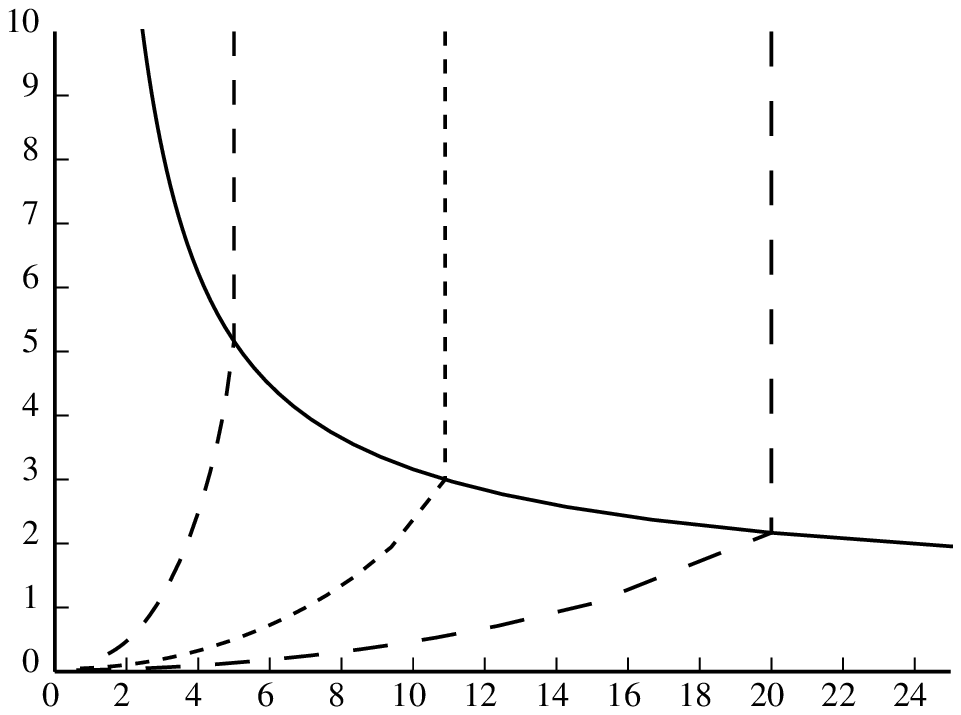,width=9cm}\end{center}}}
\put(0.5,3.2){\footnotesize $N_c$}
\put(4.5,0.15){\footnotesize $(\Lambda/\rho)^2$}
\put(3.05,5.5){\footnotesize$m_{\pi}=0$}
\put(3.05,5){\footnotesize$m_{\sigma}=2\rho_0$}
\put(5.7,1.1){\footnotesize$m_{\pi}^2=m_{\sigma}^2=\lambda\ne0$}
\put(5.2,2.3){\footnotesize $N_c^{\rm cr}$}
\end{picture}
\end{center}
\caption[]{Same plot as in Fig.~(\ref{fig1}), but for
{\em Approximation~2},
the dashed curves indicating the solutions of the two gap equations (\ref{xeqnapprox2})
and (\ref{yeqnapprox2}).
 The
 three mass values are now
 $\SigM>\rho^*$,
$\SigM<\rho^*$, $\SigM=\rho^*$, with $\rho^*=M^*\approx \sqrt{0.092}
\Lambda$, corresonding to the minimal critical
number of colors $N_c^{\rm
cr}=3$, implying a constituent quark mass above $N_c^{\rm cr}=3$
of $ 0.303\Lambda$ (short-dashed curve). The medium-dashed curve
are for a constituent quark mass $0.447\Lambda$, and the
long-dashed for $ 0.224\Lambda$.} \label{fig3}
\end{figure}

\begin{figure}[h]
\unitlength 1cm
\begin{center}
\begin{picture}(10,8)
\put(0,0){\vbox{\begin{center}
\psfig{file=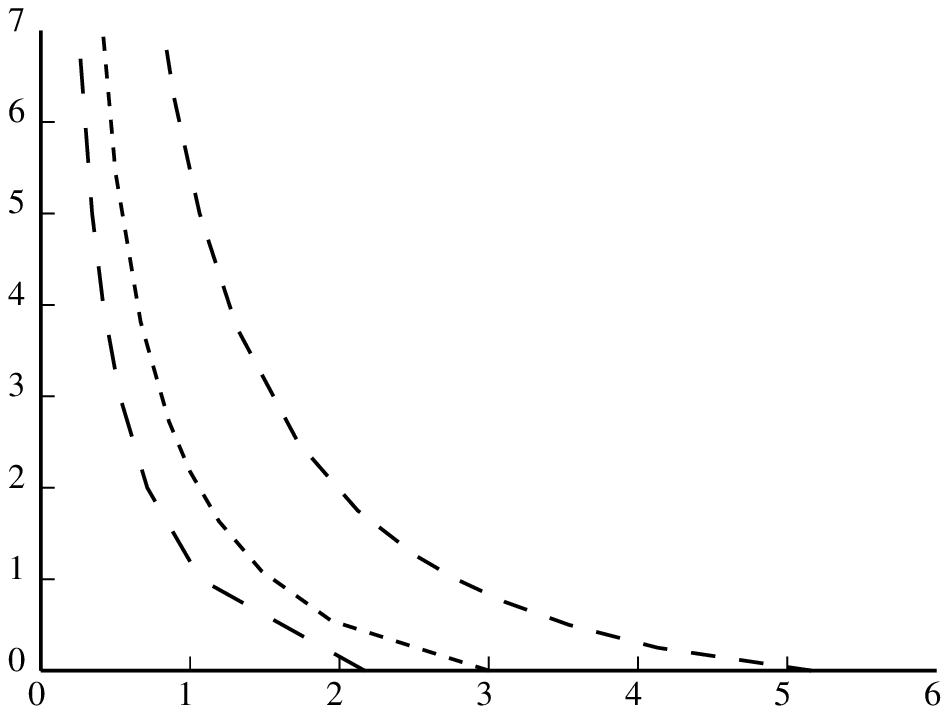,width=9cm}\end{center}}}
\put(0.5,3.2){\footnotesize $\frac{\lambda}{\rho_0^2}$}
\put(4.5,0.15){\footnotesize $N_c$}
\end{picture}
\end{center}
\caption[]{Common square masses
 $m_{\s}^2=m_{\pi}^2=\lambda$ as a function of $N_c$
in {\em Approximation~2}.
The three curves start at different critical values $N_c^{\rm cr}$
which can be read off Fig.~\ref{fig3}.}
\label{fig4}
\end{figure}

The solutions of~(\ref{xeqn})--(\ref{yeqnapprox2}) are plotted in
Figs.~\ref{fig1}--\ref{fig4}. Figs.~\ref{fig1}--\ref{fig2} are for
Appr.~1, restricted
 to the
case $\Lambda_{\pi}=\Lambda$. Qualitatively, the pictures remain
the same for different ratios $\Lambda_{\pi}/\Lambda$. Quantitatively,
there is only a shift in the critical number of color (solid curve
of Fig.~\ref{fig1}) to
 $N_c^{\rm cr}=3$ as $\Lambda_{\pi}/\Lambda$  is lowered to $0.8$,
 while it increases above the given
curve
 if $\Lambda_{\pi}/\Lambda>1$. This is due to the fact  that at the critical
 point corresponding to $\lambda=0$, one sees from Eq.~(\ref{StiffCond})
 (or from Eq.~(\ref{yeqn}) with $y=0$) that $N_c^{\rm cr}\propto$
 $(\Lambda_{\pi}/\Lambda)^2$.
 The dashed curves of
 Figs.~\ref{fig1} and~\ref{fig2} are explained in the corresponding legends.
Here we only remark that the shape of the dashed curves in
Fig.~\ref{fig1} can be understood from the gap
equations~(\ref{xeqn}) and~(\ref{yeqn}) without
 solving them, because  $x\bar{Z}(x)$ is maximal at the minimum of $N_c^{\rm cr}$.

Figures~\ref{fig3} and~\ref{fig4} correspond to Appr.~2,
in which the full momentum dependence for the pion normalization
constant is taken into account,
and
in which the pionic cutoff is
$\Lambda_{\pi}^2=4M^2$, for which we get the ratio
 $(\Lambda_{\pi}/\Lambda)^2=4x_0$. The solid curve in Fig.~\ref{fig3}
 gives the critical number of color in
 this particular case. Although the conclusion is not as strong as in
 Appr.~1, our result concerning the lack of breaking of chiral
 symmetry is robust, since the crossing with the line $N_c=3$ takes place at
a cutoff $\Lambda_{\pi}^2/M^2\gtrsim11$, which lies outside of the
admissible range $(3.3,7.3)$ implied by
 the physical value of the pion decay constant $f_{\pi}=93$ MeV,
as discussed at the
 end of the previous section.

\begin{figure}[t]
\unitlength 1cm
\begin{center}
\begin{picture}(10,8)
\put(0,0){\vbox{\begin{center}\psfig{file=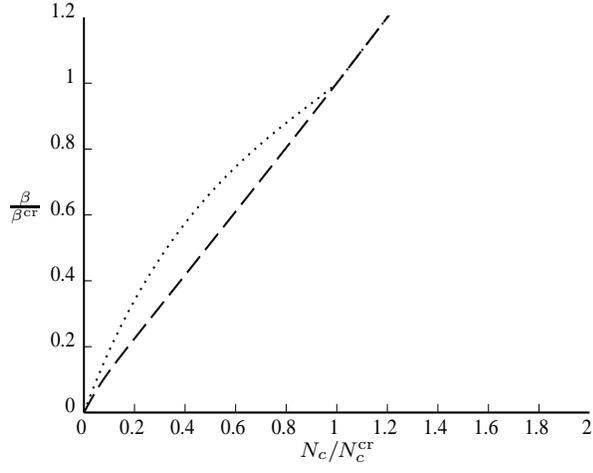,width=9cm}\end{center}}}
\put(0.5,3.5){\footnotesize $\frac{\beta}{\betacrit}$}
\put(4.38,0.25){\footnotesize $N_c/N_c^{\rm cr}$}
\end{picture}
\end{center}
\caption{{\em Approximation~1.} Reduced stiffness as a function
 of $N_c/N_c^{\rm cr}$.
The three curves below $N_c^{\rm cr}$ cannot be distinguished on
this scale. The dotted
  curve corresponds to an extra low value of $\SigM$, just to
  show that  below $N_c^{\rm cr}$ the curves deviate from a
straight line.}
\label{fig5}
\end{figure}

Finally, we give in Fig.~\ref{fig5} the stiffness as a function of
the
 number of color for Appr.~1.
The three curves depend so weakly on $ \rho _0$ that they seem to
coincide. To make the $ \rho _0$-dependence
 visible,
we have plotted an extra  dotted curve for a very small value
$\rho_0= 0.224\Lambda$ (dotted).

Let us emphasize that  these conclusions cannot be reached in the
dimensional regularization scheme since, as explained at the end of
Section~3, the integral in (\ref{CriticalStiff}) determining the
critical stiffness vanishes. Here the unphysical nature
of dimensional regularization makes its application impossible.

Before concluding, let us also remark that the  cutoff  chosen in
Appr.~2
 is completely different from that in Appr.~1, where the ratio
  of cutoffs is a constant. In Appr.~2, the ratio
of cutoffs is a
 function of $x_0$: $\Lambda_{\pi}^2/\Lambda^2=4x_0$.
 If we had taken the cutoff in the
same way as in Appr.~1 ($\Lambda_{\pi}^2/\Lambda^2=1$),
the curve giving the critical number of colors would also have had
the same shape as in Appr.~1, although the integration
would have been much more involved: the minimum number of color
would then be 5.2, whatever the value of $\Lambda^2/\rho_0^2$,  a
value which is even higher than in Appr.~1.
We see that Appr.~2 as presented above, with the physically motivated cutoff
$\Lambda_{\pi}^2=4M^2$, gives then the lowest critical number of
colors.

Our conclusions were derived from a study of only the $ \sigma $, $ \pi $
fields. The inclusion of other flavors does not help preventing
the restoration
since the associated pseudoscalar mesons are quite massive, making their
fluctuations irrelevant to the described phenomenon.

\section{Conclusion}

We have shown that within a certain nonperturbative approximation,
the Nambu--Jona-Lasinio model does not really display the
spontaneous symmetry breakdown for whose illustration it was
constructed. The fluctuations of $ \sigma $- and $\pi_a$-fields
restore chiral symmetry and make $\s$ and $\pi$ equally massive.
If our  conclusion survives
more refined approximations, this would
invalidate a large  number of publications, especially in nuclear
physics,
 which have been based on the existence of a symmetry-broken ground state of
 the model. In particular, all
studies of the temperature dependence of the
symmetry-broken state~\cite{vdb}
would deal with  nonexisting objects, thus calling for further investigations.
Finally, we note that our no-go result for the Nambu--Jona-Lasinio
 model does not imply
problems with the effective-action approach to chiral dynamics.
Certainly, there exists an effective chiral action for the meson
sector of quantum chromodynamics which does contain almost
massless pions for $N_c=3$. It is only the Nambu--Jona-Lasinio
model as it stands which is   incapable of describing these for
such a low number of colors. In fact, a recent paper
\cite{LeeOh} prompted by a first version of our preprint
points out that an extension of the Nambu--Jona-Lasinio model by
interactions involving higher-dimensional operators is not subject
to our no-go theorem. Another escape is possible by
 adding
gradient
and quartic interaction terms for $\sigma$- and $\pi$-fields to
 the
initial action, thus extending the
Nambu--Jona-Lasinio model
to  a linear sigma model \cite{klvdblin}.

\begin{acknowledgments}
We thank T. Neuhaus for discussions.
The work of B. VdB was partially supported by the Institut Interuniversitaire
 des Sciences Nucl\'eaires de Belgique and by the Alexander von Humboldt
 Foundation.
\end{acknowledgments}


\twocolumn

\end{document}